\documentclass[prd,superscriptaddress,amsfonts,amssymb,amsmath,showpacs,onecolumn]{revtex4-2}
\usepackage{bm}
\usepackage{amsfonts}
\usepackage{graphicx}
\usepackage{amsmath}
\usepackage{palatino}
\usepackage{mathpazo}
\usepackage{textcomp}
\usepackage[utf8]{inputenc}
\usepackage[T1]{fontenc}
\usepackage{float}
\usepackage{booktabs}
\usepackage{dcolumn}
\usepackage{multirow}
\usepackage{hyperref}
\hypersetup{colorlinks,citecolor=blue}
\usepackage{amsmath}
\usepackage{xcolor}
\usepackage{orcidlink}
\usepackage[caption=false]{subfig}
\usepackage{commath}
\usepackage{amssymb}

\usepackage{esdiff}
\usepackage{mathtools}
\usepackage{ifthen}
\usepackage{mathtools, amssymb} 

\usepackage{siunitx}
\usepackage{tikz}
\usetikzlibrary{hobby} 
\usetikzlibrary{arrows.meta} 
\tikzset{>={Latex[length=4,width=4]}} 
\usetikzlibrary{calc,intersections,decorations.markings}
\usepackage{siunitx}
\usepackage{xcolor} 

\colorlet{mylightblue}{blue!20}
\colorlet{myblue}{blue!50!black}
\colorlet{mydarkblue}{blue!30!black}
\colorlet{mylightred}{red!10}
\colorlet{myred}{red!50!black}
\colorlet{mydarkred}{red!60!black}
\colorlet{mydarkgreen}{green!30!black}

\tikzset{
  midarr/.style={decoration={markings,mark=at position #1 with {\arrow{stealth}}},postaction={decorate}},
  midarr/.default=0.5
}

\def\jnl@style{\it}
\def\aaref@jnl#1{{\jnl@style#1}}

\def\aaref@jnl#1{{\jnl@style#1}}

\def\aj{\aaref@jnl{AJ}}                   
\def\apj{\aaref@jnl{ApJ}}                 
\def\apjl{\aaref@jnl{ApJ}}                
\def\apjs{\aaref@jnl{ApJS}}               
\def\apss{\aaref@jnl{Ap\&SS}}             
\def\aap{\aaref@jnl{A\&A}}                
\def\aapr{\aaref@jnl{A\&A~Rev.}}          
\def\aaps{\aaref@jnl{A\&AS}}              
\def\mnras{\aaref@jnl{Mon.~Not.~Roy.~Astron.~Soc.}}             
\def\prd{\aaref@jnl{Phys.~Rev.~D}}        
\def\prc{\aaref@jnl{Phys.~Rev.~C}}  
\def\prl{\aaref@jnl{Phys.~Rev.~Lett.}}    
\def\qjras{\aaref@jnl{QJRAS}}             
\def\skytel{\aaref@jnl{S\&T}}             
\def\ssr{\aaref@jnl{Space~Sci.~Rev.}}     
\def\zap{\aaref@jnl{ZAp}}                 
\def\nat{\aaref@jnl{Nature}}              
\def\aplett{\aaref@jnl{Astrophys.~Lett.}} 
\def\apspr{\aaref@jnl{Astrophys.~Space~Phys.~Res.}} 
\def\physrep{\aaref@jnl{Phys.~Rep.}}      
\def\physscr{\aaref@jnl{Phys.~Scr}}       
\def\commat{\aaref@jnl{Comm.~Math.~Phys.}}              
\def\science{\aaref@jnl{Science}}               
\def\cqg{\aaref@jnl{Classical Quant.~Grav.}}            
\def\jpcs{\aaref@jnl{JPCS}}                                     
\def\ijmpd{\aaref@jnl{Int.~J.~Mod.~Phys.~D}}                    
\def\grg{\aaref@jnl{Gen.~Relat.~Gravit.}}               
\def\rpp{\aaref@jnl{Rep.~Prog.~Phys.}}          
\def\npa{\aaref@jnl{Nucl.~Phys.~A}}        
\def\lrr{\aaref@jnl{Living Rev.~Rel.}}                   
\def\jcap{\aaref@jnl{J.~Cosmology Astropart.~Phys.}}    
\def\rmp{\aaref@jnl{Rev.~Mod.~Phys.}}   
\def\epjc{\aaref@jnl{Eur.~Phys.~J.~C}}

\hypersetup{linkcolor=blue}

\allowdisplaybreaks[1]

\begin{document}

\color{black}       

\title{Probing the existence of the ZTF Casimir wormholes in the framework of $f(\mathcal{R})$ gravity}

\author{Oleksii Sokoliuk\orcidlink{0000-0003-4503-7272}}
\email{oleksii.sokoliuk@mao.kiev.ua}
\affiliation{Main Astronomical Observatory of the NAS of Ukraine (MAO NASU),\\
Kyiv, 03143, Ukraine}
\affiliation{Astronomical Observatory, Taras Shevchenko National University of Kyiv, \\
3 Observatorna
St., 04053 Kyiv, Ukraine}

\author{Alexander Baransky\orcidlink{0000-0002-9808-1990}}
\email{abransky@ukr.net}
\affiliation{Astronomical Observatory, Taras Shevchenko National University of Kyiv, \\
3 Observatorna
St., 04053 Kyiv, Ukraine}
\author{P.K. Sahoo\orcidlink{0000-0003-2130-8832}}
\email{pksahoo@hyderabad.bits-pilani.ac.in}
\affiliation{Department of Mathematics, Birla Institute of Technology and
Science-Pilani,\\ Hyderabad Campus, Hyderabad-500078, India.}

\date{\today}
\begin{abstract}

For the spherically symmetric static traversable wormholes, supported by the Casimir energy in $f(\mathcal{R})=\mathcal{R}+\alpha \mathcal{R}^2$ Quadratic, $f(\mathcal{R})=f_0 \mathcal{R}^n$ power-law Modified Gravity (MOG) theories we investigate energy conditions and dynamical stability of the wormhole solutions. Especially, we study Zero Tidal Forces (ZTF) Casimir WH's with anisotropic fluid located at the throat.  By using the Casimir energy density and modified Einstein Field Equations (EFE's) we derived suitable shape functions for each modified gravity of our consideration. The stability of Casimir traversable wormholes in different modified gravity theories is also analyzed in our paper with a modified Tolman-Oppenheimer-Voklov (MTOV) equation. Besides, we have numerically solved MTOV and derived hydrodynamical, anisotropic and extra forces, that is present due to the non-conserved stress-energy tensor. Moreover, other fundamental quantities, such as Volume Integral Quantifier and total gravitational energy were derived.

\end{abstract}

\maketitle

\section{Introduction} \label{sec:1}
Over the decades, a large number of works have been written that have been devoted to the study of wormholes (see \cite{Ellis1973,Bronnikov1973,Morris1988,Hochberg1993,Visser1989,Visser1997,Kim2001,Dadhich2002,Kuhfittig2003}). Wormholes, in fact are just a connection between two points at the four-dimensional manifold (our spacetime), two universes or even branes (then, wormhole throat is located at the bulk). In the General theory of Relativity (here and further - just GR),  wormholes are just the solution of Einstein field equations \cite{Capozziello2011}. Because of that wormholes could allow the observer to travel quickly over long distances not exceeding the speed of light, this field of study become popular over the last years. First wormhole solution was presented by \cite{Einstein1935} and it was called Einstein-Rosen bridge. This solution connects two universes and occurs in the maximally-extended Schwarzschild spacetime. Because of that fact, there is always present the event horizon, and thus, anyone trying to escape wormhole throat will fall onto the singularity. On the other hand, five decades after the discovery of the Einstein-Rosen Bridge, Morris and Thorne found another solution that could theoretically allow an outside observer to travel freely through the wormhole \cite{Ellis1973}. But, as it turned out, in GR to hold necessary condition $\tau_0>\rho_0c^2$ (where $\tau_0$ is a radial tension) there always must be present so-called exotic fluid at the throat, which violates Null Energy Condition (NEC). To overpass this problem (or at least minimize the amount of NEC violating matter at the throat), one could assume modified Einstein-Hilbert action integral (i.e. modified gravity).
\subsection{Modified gravity}
General relativity is known to describe the universe well in general, but recent cosmological observations require modifying the gravity of classical relativity $\mathcal{R}$ into some alternative that could describe well, for example, the late-time acceleration phase of the universe \cite{Komatsu2011, Perlmutter1999,Riess2007,Riess1998,Suzuki2012}, and also there exist one form of the $f(\mathcal{R})$ function that could describe inflation as well as the late time acceleration without any additional fields \cite{Odintsov2017}. There was many attempts to modify GR, One of the most common types of modified gravity is the Ricci scalar function $f(\mathcal{R})$, which replace classical Ricci scalar gravity $\mathcal{R}$ and modifies Einstein-Hilbert action integral. This theory was originally proposed in \cite{Buchdahl1970}. It has drawn the attention of cosmologists, because it can
provide a geometric mechanism for the description of inflation \cite{Brooker2016,Huang2013,Starobinsky1980} and solve the dark energy problem \cite{Capozziello2011,Nojiri2017}. In the traversable wormhole field of study there was written big amount of papers with the various forms of arbitrary function $f(\mathcal{R})$, also some works present with considered $f(\mathcal{R},T)$ formalism.

Also, on the topic of different static/rotating wormhole spacetimes, supported by the Casimir energy (nor by the GUP corrected Casimir energy) there was written several papers, such as \cite{Jusufi2020,Muniz2021,Santos2021,Sorge2020}. Moreover, paper \cite{https://doi.org/10.48550/arxiv.2107.11375} investigate spherically symmetric wormholes in the presence of tidal forces and electromagnetic field sourced by the both Casimir and GUP corrected Casimir effects, in the pioneering work of \cite{Garattini2021} authors introduce ZTF Casimir wormholes with the Yukawa potential corrections. Within the present study, we will focus on the investigations of the Morris-Thorne wormholes sourced by the Casimir effect with the Zero Tidal Forces (ZTF) present. Such assumption has been made for the sake of simplicity and the viability of the wormhole solution (in \cite{Morris1988}, it was prescribed that in order for wormhole to be humanly traversable, tidal forces must be bearably small).
\subsection{Casimir effect}
The Casimir effect appears between two planes parallel, closely spaced, uncharged, metallic plates in vacuum \cite{Garattini2019}. The main interesting property of Casimir effect is attractive force that arise between two parallel planes. This force is present because of the renormalized energy in form:
\begin{equation}
    E^{\mathrm{Ren}} (a) = -\frac{\hslash c \pi^2S}{720a^3},
\end{equation}
where $S$ is the surface of the parallel planes and $a$ is separation between them. Consequently, one could derive general view of the attractive force in the following form:
\begin{equation}
    F(a) = -\frac{\mathrm{d}E^{\mathrm{Ren}(a)}}{\mathrm{d}a} = -3\frac{\hslash c \pi^2 S}{720 a^4}
\end{equation}
Finally, we also could derive suitable pressure and energy density:
\begin{equation}
    P(a) = \frac{F(a)}{S} = -3\frac{\hslash c \pi^2}{720 a^4}
\end{equation}
\begin{equation}
    \rho_C (a) = -\frac{\hslash c \pi^2}{720 a^4}
    \label{eq:1.4}
\end{equation}
It is obvious that with given pressure and energy density we could assume Equation of State (further - EoS) $P=\omega\rho_C$ with EoS parameter $\omega=3$ \cite{Garattini2019}. In our article, we want to replace exotic matter with Casimir-like matter (which has Casimir energy density). In order to do so, one could assume the proper transformation of the plate separation $a\to r$. It was found in the pioneering works of \cite{Morris1988,Garattini2019}, that using such replacement could keep the energy-momentum tensor conservation, i.e. the physical viability of the theory (however, in some cases $\nabla^\mu T_{\mu\nu}=0$ could not be obeyed, as expected).
\subsection{Article organisation}
Article is organised as follows: in the Section (\ref{sec:1}) we provide the general introduction into the topic of traversable wormholes, Casimir effect and modified gravity. In the Section (\ref{sec:2}) we present the energy conditions, that we consider in the paper, spherically symmetric static wormhole spacetime and EFE's in the classical GR gravity. Further, in the Section (\ref{sec:3}) we modify Einstein-Hilbert action and present $f(\mathcal{R})$ gravity formalism, derive suitable shape functions with different forms of the function $f(\mathcal{R})$. Moreover, in the Section (\ref{sec:4}) we probe the stability of the Casimir WH anisotropic matter in different kinds of $f(\mathcal{R})$ gravity. In the Section (\ref{sec:5}), we analyze energy conditions of the Casimir wormholes. In the next Section, namely Section (\ref{sec:66}) we derive the exact volume of NEC violating matter content that is located nearby the wormhole throat for both WH models of our consideration. On the other hand, in the Section (\ref{sec:7}) we introduce new term of interest, namely total gravitational energy and investigate this term in details within the spherically symmetric wormhole spacetimes. Finally, we present main conclusions in the Section (\ref{sec:6}).
\section{Casimir wormholes in the classical GR gravity} \label{sec:2}
\subsection{Energy conditions}
As you know, one of the most possible variants of the existing traversable wormhole is the Morris-Thorne wormhole. But in the classic GR, the existence of such objects requires exotic matter located at the throat of the wormhole. Casimir Energy is currently the only possible substitute for exotic matter, that could be obtained in the laboratory. Firstly, we could present energy conditions, that we will use in this paper \cite{Sokoliuk:2021rtv}:
\begin{itemize}
    \item Null Energy Condition (NEC): $\rho+p_r \geq 0\land \rho+p_t \geq 0$
    \item Weak Energy Condition (WEC): $\rho\geq0$ and $\rho+p_r \geq 0 \land \rho+p_t \geq 0$
    \item Strong Energy Condition (SEC): $\rho + p_r + 2p_t \geq 0$
    \item Dominant Energy Condition (DEC): $\rho \geq |p_r| \land \rho \geq |p_t|$
\end{itemize}
In the classical GR gravity, exotic matter at the throat violates NEC.
\subsection{Einstein Field Equations}
Because of the quantum nature of the Casimir effect, one could want to replace classical EFE to the semi-classical one (we use unit system with $\kappa=1$) \cite{Hochberg1996}:
\begin{equation}
    G_{\mu\nu} + \Lambda g_{\mu\nu} =  \langle T_{\mu\nu}\rangle^{\mathrm{Ren}}.
\end{equation}
Here $\langle T_{\mu\nu}\rangle^{\mathrm{Ren}}$ is renormalized expectation value of the stress-energy tensor, which describes matter fields contribution. Also, we assume that cosmological constant is zero (i.e. $\Lambda=0$).
Then, we could define spherically symmetric and static wormhole spacetime as follows \cite{Morris1988}:
\begin{equation}
\begin{gathered}
    ds^2 = -e^{2\Omega(r)}dt^2+\frac{1}{1-\dfrac{b(r)}{r}}dr^2+r^2d\theta^2 + r^2\sin^2\theta d\phi^2,
\label{eq:1}
\end{gathered}
\end{equation}
where $\Omega(r)$ is so-called redshift function and $b(r)$ is shape function. Shape function must always obey flaring out condition $\frac{b(r)-rb'(r)}{b(r)^2}>0$, as well as the conditions $b'(r_0)<1$ and $b(r_0)=r_0$ at the wormhole throat. In turn, we assume that our spacetime has no horizons, and thus redshift factor must be finite everywhere \cite{Garattini2019}, and also tidal forces must be relatively small \cite{Morris1988}. From this conditions we have chosen the case with zero tidal forces (i.e. constant redshift factor). With the given metric tensor (\ref{eq:1}) we could rewrite EFE's \cite{Oliveira2021}:
\begin{equation}
    \rho = \frac{b'}{r^2}
\end{equation}
\begin{equation}
    p_r = -\frac{b}{r^3}+2\bigg(1-\frac{b}{r}\bigg)\frac{\Omega'}{r}
\end{equation}
\begin{equation}
   p_t =  \bigg(1-\frac{b}{r}\bigg)\bigg[\Omega''+(\Omega')^2-\frac{rb'-b}{2r^2(1-b/r)}\Omega'-\frac{rb'-b}{2r^3(1-b/r)}+\frac{\Omega'}{r}\bigg]
\end{equation}
Where $\rho$ is obviously energy density, $p_r$ is radial pressure and $p_t$ is tangential one (for the isotropic WH fluid we have that $p_r=p_t$). In the further investigation, we will use the constant redshift function (therefore, ZTF wormhole solution) because of the beforehand mentioned reasons.
\section{$f(\mathcal{R})$ gravity} \label{sec:3}
In $f(\mathcal{R})$ theory of gravity the EH Action Integral is modified as follows \cite{Buchdahl:1983zz}:
\begin{equation}
\begin{gathered}
    \mathcal{S}_\mathcal{R}=\frac{1}{2}\int_{\mathcal{M}} (\mathcal{R}+\mathcal{L}_M) \sqrt{-g} d^4x \\
    \Downarrow\\
    \mathcal{S}_{f(\mathcal{R})}=\frac{1}{2}\int_{\mathcal{M}} [f(\mathcal{R})+\mathcal{L}_M)]\sqrt{-g} d^4x,
\end{gathered}
\end{equation}
where $\mathcal{L}_\textrm{M}$ is the Lagrangian of the matter fields. Then, by varying the Einstein-Hilbert action we could obtain (modified) EFE \cite{Sokoliuk:2021rtv}:
\begin{equation}
\begin{gathered}
    G^{(0)}_{\mu\nu}\equiv R_{\mu\nu}-\frac{1}{2}g_{\mu\nu}\mathcal{R}=\frac{T^\textrm{M}_{\mu\nu}}{f'(\mathcal{R})}+g_{\mu\nu}\frac{[f(\mathcal{R})-\mathcal{R}f'(\mathcal{R})]}{2f'(\mathcal{R})}\\ + \frac{[\nabla_\mu\nabla_\nu f'(\mathcal{R})-g_{\mu\nu}\Box f'(\mathcal{R})]}{f'(\mathcal{R})}.
\end{gathered}
\label{eq:13}
\end{equation}
Hence, we have that \cite{Lobo2009}:
\begin{equation}
    \rho = \frac{f_R b'}{r^2}
    \label{eq:3.3}
\end{equation}
\begin{equation}
    p_r = -\frac{b f_R}{r^3}+\frac{f_R'}{2r^2}(b'r-b)-f_R''\bigg(1-\frac{b}{r}\bigg)
    \label{eq:3.4}
\end{equation}
\begin{equation}
   p_t =  -\frac{f_R'}{r}\bigg(1-\frac{b}{r}\bigg)+\frac{f_R}{2r^3}(b-b'r)
   \label{eq:3.5}
\end{equation}
where $f_R=df/d\mathcal{R}$. Further we will derive suitable shape function with specified arbitrary function $f(\mathcal{R})$.
\subsection{Quadratic gravity}
Quadratic gravity is well-known first Starobinsky inflationary model. In this kind of gravity we have following $f(\mathcal{R})$ \cite{Arapo2017}:
\begin{equation}
    f(\mathcal{R}) = \mathcal{R} + \alpha \mathcal{R}^2
\end{equation}
Using Casimir energy density from Equation (\ref{eq:1.4}) we could obtain suitable shape function for our Casimir wormhole in the chosen MOG (we replaced plate separation $a$ by the radial coordinate $r$). In order to do so, we will solve the field equations (\ref{eq:3.3}), (\ref{eq:3.4}) and (\ref{eq:3.5}) that could be converted into the
\begin{equation}
    -\frac{\hslash \pi^2}{720r^4}=\frac{(1+2\alpha \mathcal{R})b'}{r^2}
\end{equation}
analytically using the \texttt{Mathematica} ODE solver \texttt{NDSolve}:
\begin{equation}
\begin{gathered}
    b(r) = c_1+\frac{r \sqrt{9 r^4-\frac{1}{5} \pi ^2 \alpha  \hslash}}{72 \alpha }+\Bigg[i \pi ^{3/2} \hslash \sqrt{\pi ^2-\frac{45 r^4}{\alpha  \hslash}}
    F\left(i \sinh ^{-1}\left(\frac{\sqrt{3} \sqrt[4]{5} r \sqrt{-\frac{1}{\sqrt{\hslash} \sqrt{\alpha }}}}{\sqrt{\pi
   }}\right)-1\right)\Bigg]\\
   \Bigg/\Bigg[36\ 5^{3/4} \sqrt{-\frac{1}{\sqrt{\alpha } \sqrt{\hslash}}} \sqrt{135 r^4-3 \pi ^2 \alpha  \hslash}\bigg]-\frac{r^3}{24 \alpha }
   \label{eq:3.6}
   \end{gathered}
\end{equation}
Therefore, constant of integration could be derived by imposing the so-called throat condition $b(r_0)=r_0$:
\begin{equation}
\begin{gathered}
    c_1 = \frac{-r_0 \sqrt{9 r_0^4-\frac{1}{5} \pi ^2 \alpha  \hslash}+3 r_0^3+72 \alpha  r_0}{72 \alpha }\\
    -\frac{i \pi ^{3/2} \hslash \sqrt{\pi ^2-\frac{45 r_0^4}{\alpha  \hslash}} F\left(i \sinh ^{-1}\left(\frac{\sqrt{3} \sqrt[4]{5} r_0 \sqrt{-\frac{1}{\sqrt{\hslash} \sqrt{\alpha }}}}{\sqrt{\pi
   }}\right)-1\right)}{36\ 5^{3/4} \sqrt{-\frac{1}{\sqrt{\alpha } \sqrt{\hslash}}} \sqrt{135 r_0^4-3 \pi ^2 \alpha  \hslash}}
   \end{gathered}
\end{equation}
Now, since we already derived all of the necessary quantities, we could proceed further and probe our shape function in terms of physical viability.
\begin{figure*}[!htbp]
    \centering
    \includegraphics[width=\textwidth]{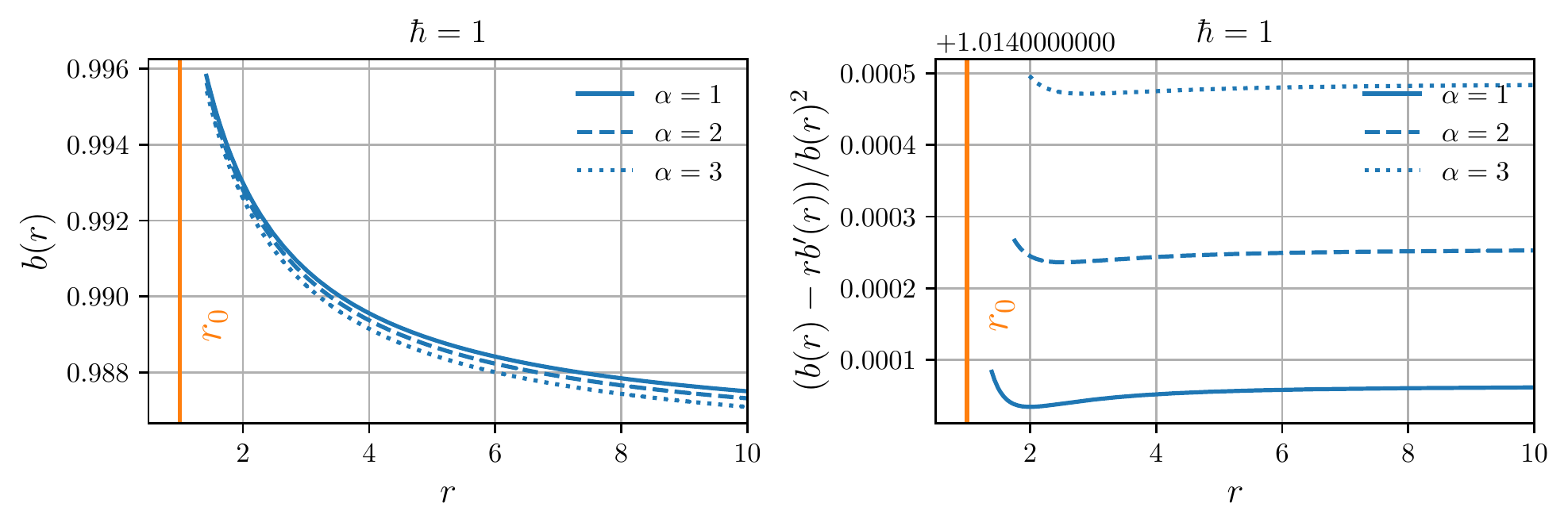}
    \caption{Shape function and flare-out condition for Quadratic gravity with throat radius $r_0=2$}
    \label{fig:1}
\end{figure*}
\newline On the Figure (\ref{fig:1}) we numerically solved Equation (\ref{eq:3.6}) with varying values of free MOG parameter $\alpha$. As one could notice, generally flare-out condition is satisfied for a given Casimir WH shape function with relatively small values of $\alpha$.

\subsection{Power-law gravity}
\begin{figure*}
    \centering
    \includegraphics[width=\textwidth]{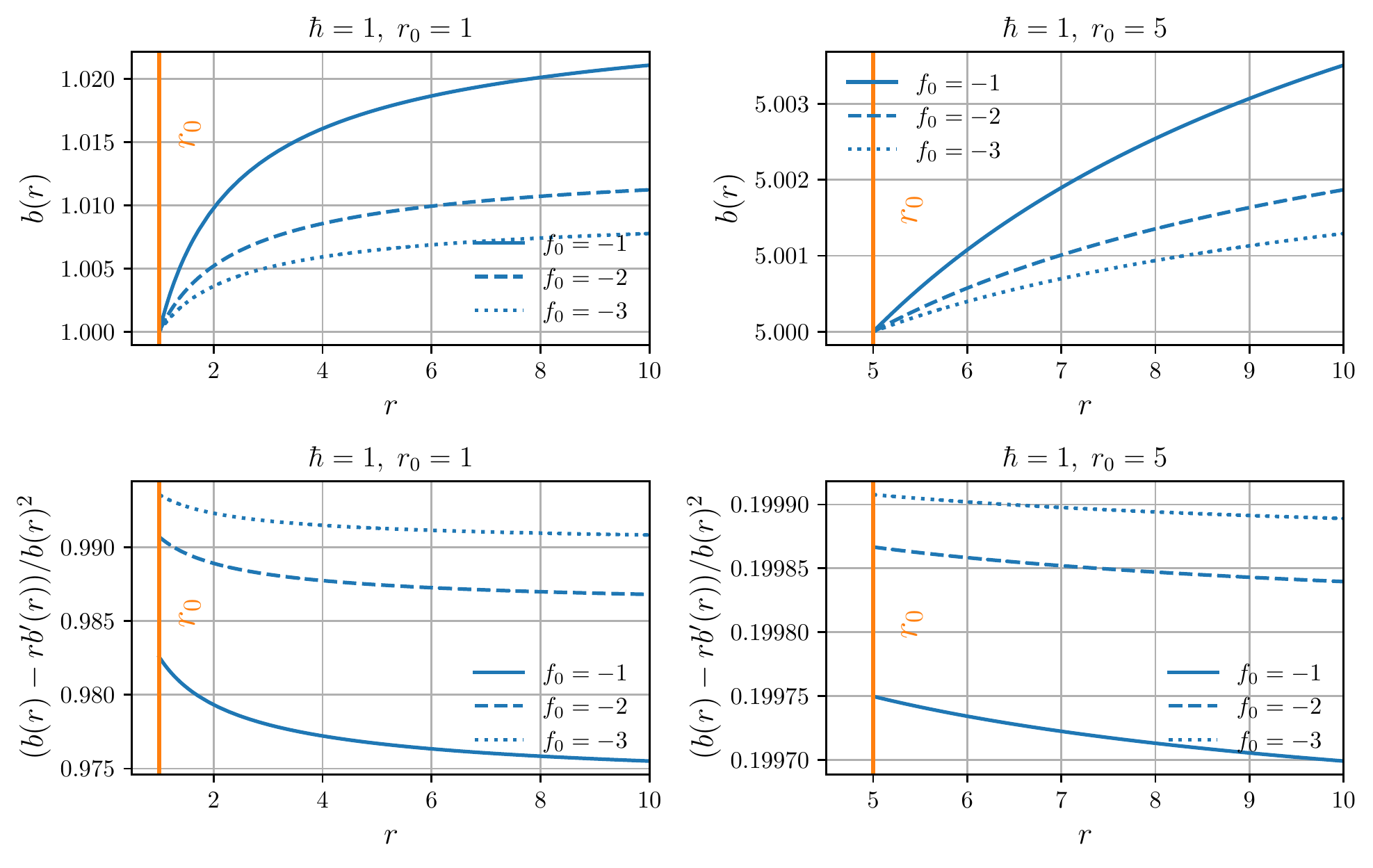}
    \caption{Shape function and flare-out condition for near GR ($n=1.1$) power-law gravity with throat radius $r_0=2$}
    \label{fig:3}
\end{figure*}

In this subsection we consider following form of the $f(\mathcal{R})$ function:
\begin{equation}
    f(\mathcal{R}) = f_0 \mathcal{R}^n
\end{equation}
This modified gravity was firstly described in the papers \cite{Carloni2005,Nojiri2006}. Thus, we could derive suitable Casimir WH shape function from (modified) EFE's with the use of Casimir energy density matching:
\begin{equation}
    -\frac{\hslash \pi^2}{720r^4}=\frac{(f_0n \mathcal{R}^{n-1})b'}{r^2}
\end{equation}
Consequently, analytical solution to the aforementioned equation could be found:
\begin{equation}
    b(r) = c_1+\frac{45^{-1/n} n \pi ^{2/n} r^3 \left(-\frac{\hslash 2^{-n-3}}{f_0 n r^4}\right)^{\frac{1}{n}}}{3 n-4}
    \label{eq:18}
\end{equation}
Following the same procedure that was applied for the quadratic $f(\mathcal{R})$ gravitation, one may derive constant $c_1$ by imposing the well-known throat condition $b(r_0)=r_0$:
\begin{equation}
    c_1 = \frac{45^{-1/n} n \pi ^{2/n} r_0^3 \left(-\frac{\hslash 2^{-n-3}}{f_0 n r_0^4}\right)^{\frac{1}{n}}}{4-3 n}+r_0
\end{equation}
We numerically solved Equation (\ref{eq:18}) on  the Figure (\ref{fig:3}) with different values of throat radius and free parameter $f_0$. Flare-out condition is violated for relatively big values of throat radius and $n$. Also, we could conclude that $n$ should have values near to the $n=1$ (GR case), judging by the experimental data from \cite{Clifton2005}. Also, there is no positive $f_0$ solutions for the shape function in the power-law gravity.
\section{Casimir wormhole stability}  \label{sec:4}
\label{sec:4}
In this section we will analyze stability of the Casimir wormhole anisotropic matter. For this goal we have so-called Tolman-Oppenheimer-Volkov equation, which in modified gravity reads \cite{Oppenheimer1939,Poncede1993,Rahaman2014,Tolman1939}:
\begin{equation}
\begin{gathered}
    \frac{dp_r}{dr}+\underbrace{\Omega(r)'(\rho+p_r)}_\text{if $\Omega=0,0$}+\frac{2}{r}(p_r-p_t) + F_{\mathrm{ex}}=0\\
    \Downarrow\\
    \frac{dp_r}{dr}+\frac{2}{r}(p_r-p_t) + F_{\mathrm{ex}} =0,
\end{gathered}
\label{eq:45}
\end{equation}
where extra force $F_{\mathrm{ex}}$ arise because of the discontinuous stress-energy tensor in $f(\mathcal{R})$ gravity. In the next subsections, we will investigate the wormhole stability in the various modified gravities of our consideration.
\subsection{Quadratic gravity}
Extra force for quadratic (Starobinsky-like) gravitation will be derived numerically, since the expression is very complex. Consequently, numerical solution for the MTOV equilibrium problem for quadratic gravity is located on the first row of Figure (\ref{fig:44}). As one could easily notice, only arbitrary small extra force is needed in order to keep traversable wormhole stable within the Starobinky $f(\mathcal{R})$ gravity. Besides, hydrodynamical and extra forces $F_{\mathrm{H}}$, $F_{\mathrm{ex}}$ show positive behavior on the whole manifold and anisotropic force $F_{\mathrm{A}}$ shows negative behavior respectively.
\subsection{Power-law gravity}
Beforehand mentioned extra force could be therefore derived analytically for power-law $f(\mathcal{R})$ gravity:
\begin{equation}
\begin{gathered}
F_{\mathrm{ex}}=\bigg(f_0 2^n 45^{-1/n} (n-1) \left(45^{-1/n} \pi ^{2/n} \left(-\frac{\hslash
   2^{-n-3}}{f_0 n r^4}\right)^{\frac{1}{n}}\right)^{n-1} \left(2 n^2 (5 n-4)
   \pi ^{2/n} r^3 \left(-\frac{\hslash 2^{-n-3}}{f_0 n r^4}\right)^{\frac{1}{n}}\right)\\
   +(7
   n-8) (7 n-4) r_0 \left(45^{\frac{1}{n}} (3 n-4)-n \pi ^{2/n} r_0^2
   \left(-\frac{\hslash 2^{-n-3}}{f_0 n r_0^4}\right)^{\frac{1}{n}}\right)-4\times
   45^{\frac{1}{n}} (4-3 n)^2 (3 n-2) r\bigg)\\
   \bigg/\bigg(n^2 (3 n-4) r^4\bigg)
\end{gathered}
\end{equation}

\begin{figure*}[!htbp]
    \centering
    \includegraphics[width=\textwidth]{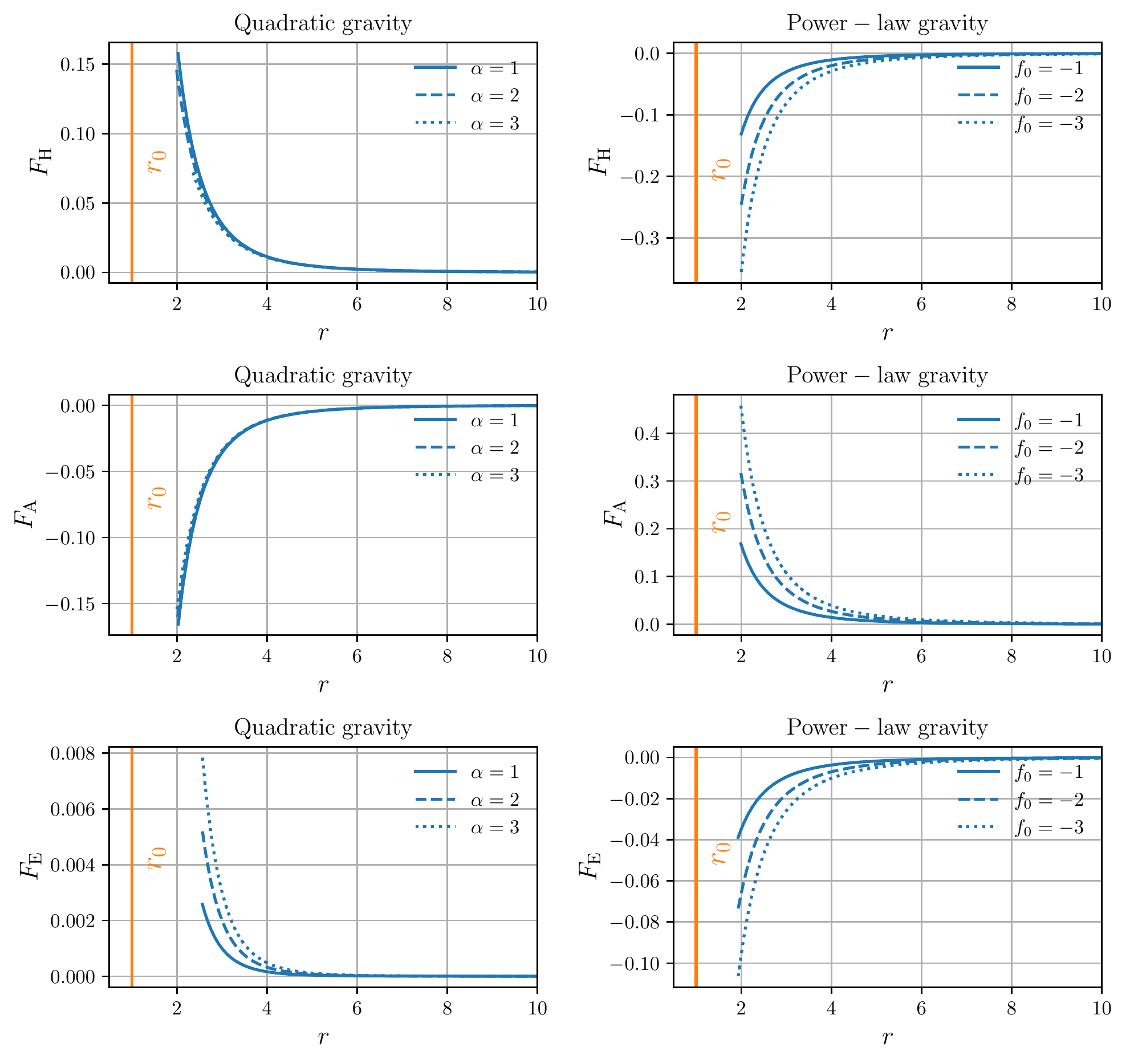}
    \caption{Hydrodynamical, anisotropic forces for quadratic (left plot) and power-law (right one) gravity with throat radius $r_0=1$ and $n=1.1$}
    \label{fig:44}
\end{figure*}

On the second row of Figure (\ref{fig:44}) we placed numerical solutions for the present in MTOV hydrodynamical, anisotropic forces within the power-law MOG of our consideration. As it was remarked during the numerical analysis, extra force has smallest contribution to the MTOV if $f_0\to0$, as expected. In relation to the Quadratic gravity case, difference between hydrodynamical and anisotropic forces with the growing values of $f_0$ is bigger and therefore extra force grows with $f_0\to-\infty$ (which signifies the deviation from GR). It is also worth to notice that $F_{\mathrm{H}}\land F_{\mathrm{ex}}\leq0$ and $F_{\mathrm{A}}\geq0$ generally.
\section{Energy conditions within $f(\mathcal{R})$ gravitation} \label{sec:5}
In this section and following subsections, we as usual going to apply the various energy conditions, such as Null, Weak, Dominant and Strong Energy Conditions for the quadratic and power-law $f(\mathcal{R})$ gravities.
\subsection{Quadratic gravity}
\begin{figure*}[!htbp]
    \centering
    \includegraphics[width=\textwidth]{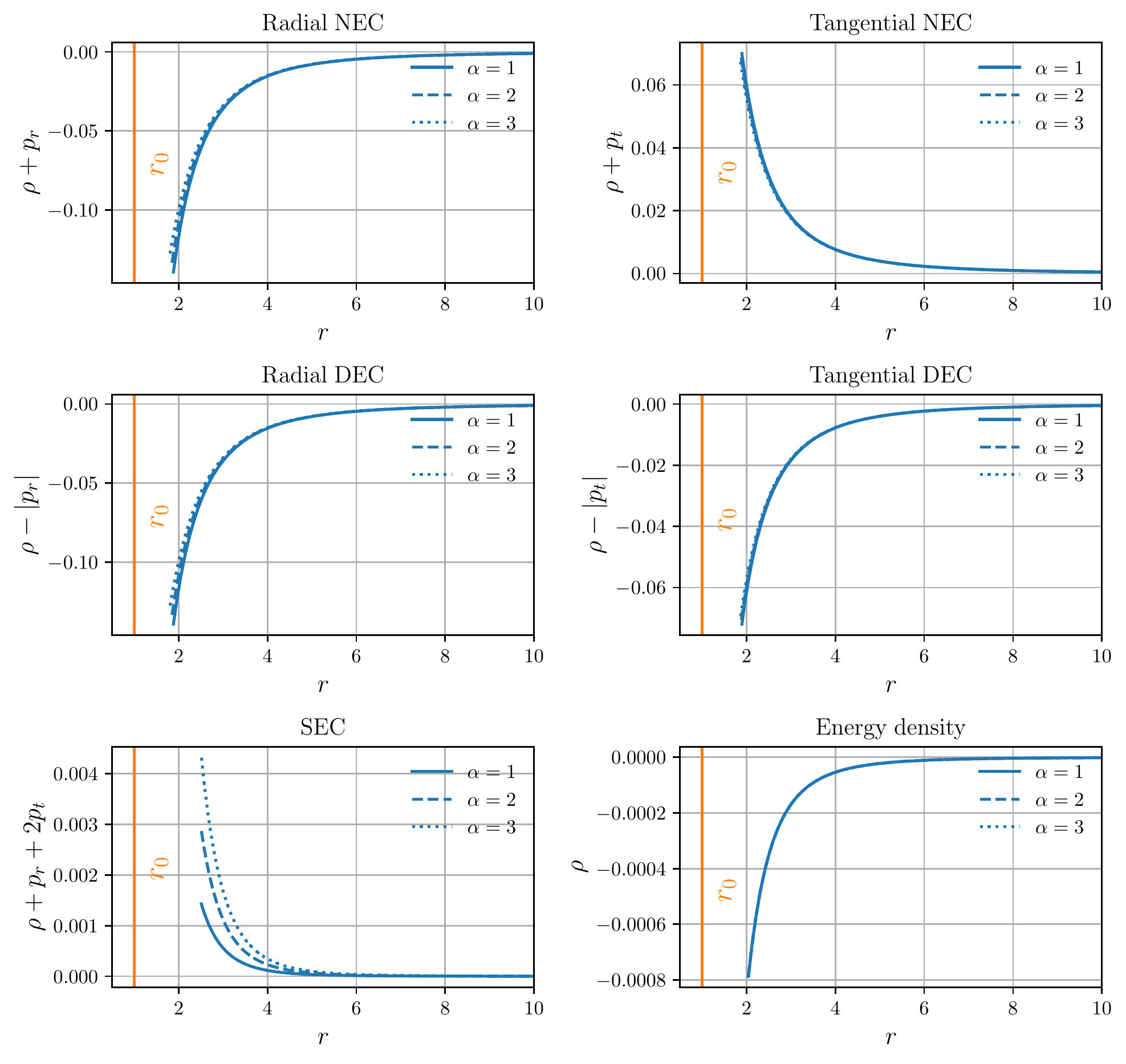}
    \caption{Null, Dominant and Strong energy conditions for Quadratic gravity with throat radius $r_0=1$}
    \label{fig:2}
\end{figure*}

On the Figure (\ref{fig:2}) we depicted and probed energy conditions of the Casimir wormhole in the Quadratic gravity with varying values of the additional degree of freedom $\alpha$. As we noticed, generally NEC is violated for radial pressure and validated for the tangential one. In turn, dominant energy condition was violated for both radial pressure tangential pressure for any value of $\alpha$. SEC was also violated at the wormhole throat. On the last, sixth plot of the Figure (\ref{fig:2}) we additionally plot the energy density. As it was unveiled, energy density $\rho$ is unvariant under the variation of $\alpha$ and it is negative, therefore both radial and tangential WEC are violated.
\subsection{Power-law gravity}
\begin{figure*}[!htbp]
    \centering
    \includegraphics[width=\textwidth]{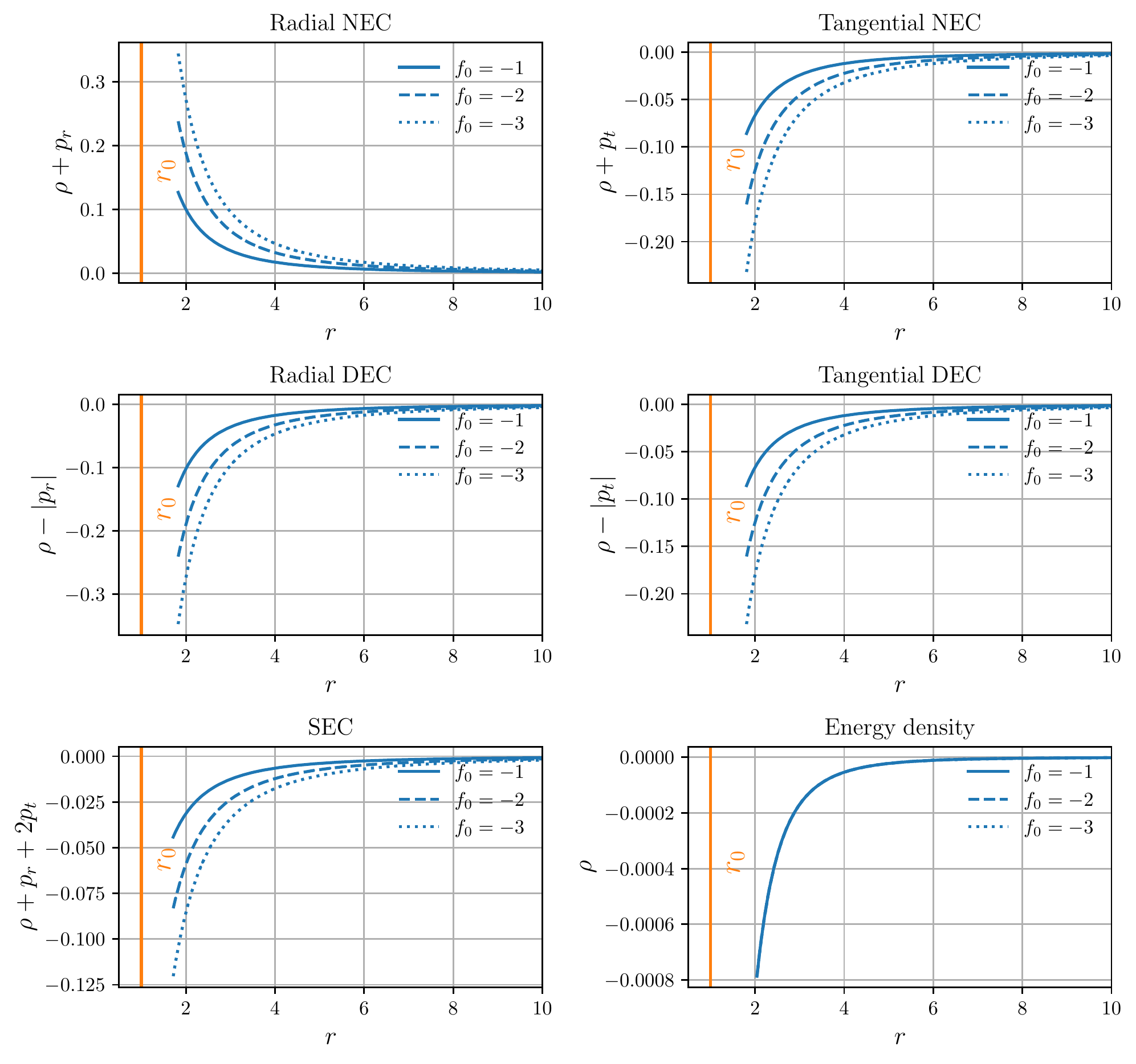}
    \caption{Null, Dominant and Strong energy conditions, energy density for near GR ($n=1.1$) power-law gravity with throat radius $r_0=1$}
    \label{fig:4}
\end{figure*}

Again, on the Figure (\ref{fig:4}) we illustrated energy conditions for Casimir wormhole in the power-law gravity with near GR value of $n$. During the numerical analysis we noticed that in general NEC was validated for radial pressure and violated for tangential one, DEC was violated for radial and tangential pressures (even on the bigger values of $n$). Finally, SEC was validated in relation to the Quadratic gravitation and judging by the negative energy density, WEC was violated. However, if $f_0$ will be sufficiently small and negative, some of the aforementioned energy conditions could be validated near the wormhole throat, which signifies the absence of exotic matter.
\section{Volume Integral Quantifier (VIQ)}\label{sec:66}
\begin{figure}[!htbp]
    \centering
    \includegraphics[width=\textwidth]{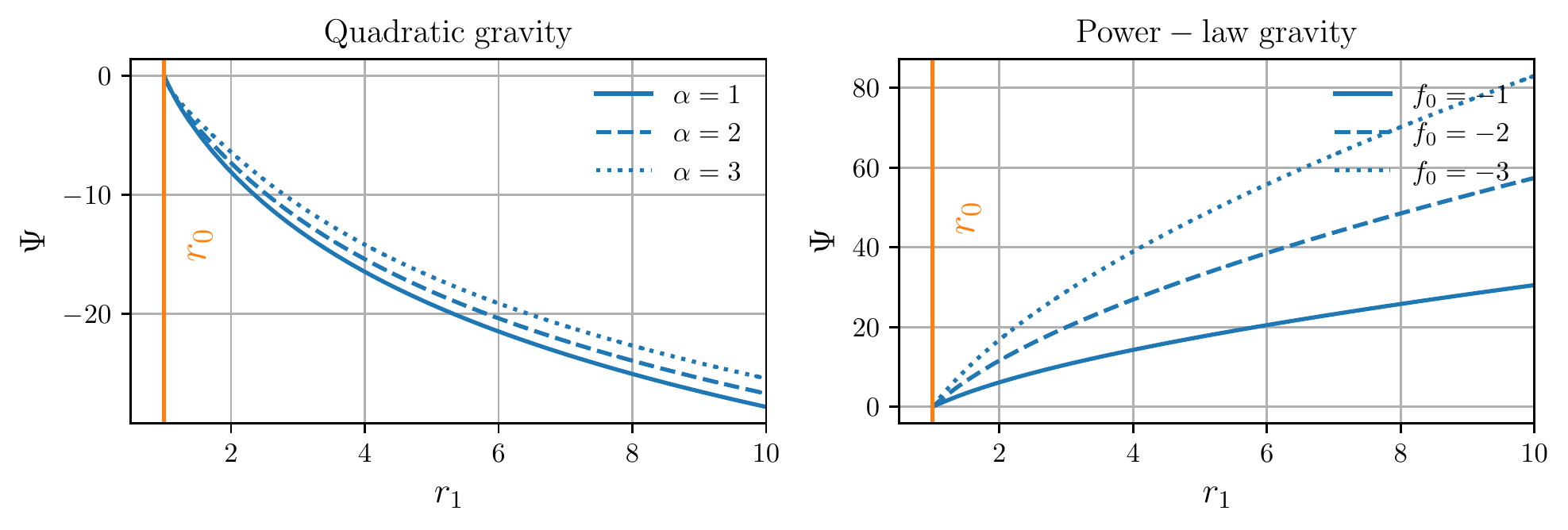}
    \caption{Volume Integral Quantifier (VIQ) for (\textit{first plot}) Quadratic gravity and (\textit{second plot}) power-law gravity with $n=1.1$}
    \label{fig:VIQ}
\end{figure}

Volume Integral Quantifier (further - VIQ) is widely used for the derivation of the total exotic matter (matter, that violates Average Null
Energy Condition (ANEC)) amount nearby the wormhole throat. Within the spacetime, imposing spherical symmetry, we could write down VIQ in it's usual form \cite{2019MPLA...3450303S}:
\begin{equation}
    \Psi = \int^\infty_{l_0}\int^\pi_0\int^{2\pi}_0[\rho+p_r]\sqrt{-g}drd\theta d\phi
    \label{eq:24}
\end{equation}
Simplifying above equation, after some tedious algebra one could came up with the cuvilinear integral solution, which looks exactly like the expression below:
\begin{equation}
    \Psi = \oint [\rho+p_r]dV=2\int^\infty_{r_0}[\rho+p_r]4\pi r^2 dr
\end{equation}
As we see, in the equation above integration bounds are infinity, and as it was found in the \cite{PhysRevD.87.084030}, for a wormhole to be asymptotically flat, VIQ with infinite bounds over radial coordinate must diverge. Therefore, it will be handful to present a energy-momentum tensor cut-off scale at some point $r_1$:
\begin{equation}
    \Psi = 2\int^{r_1}_{r_0}[\rho+p_r]4\pi r^2 dr
    \label{eq:6.3}
\end{equation}
We numerically integrate Equation (\ref{eq:6.3}) and derive VIQ for both quadratic and power-law gravities and graphically represent the solutions on the Figure (\ref{fig:VIQ}). As expected, with the growing value of cutoff value of $|\Psi|\to\infty$.

\section{Total gravitational energy}\label{sec:7}
The last feature that we are going to investigate in this paper is the so-called total gravitational energy. It is known that the value of total gravitational energy must be negative in order for the matter to be composed from the baryonic particles \cite{Rahaman:2013ywa}. This fundamental quantity was firstly presented in the pioneering work of Lyden-Bell et al. \cite{Lynden-Bell:2007gof} in the absence of black holes for stationary spacetimes (which is exactly what we are looking for). In the aforementioned work, total gravitational energy were composed of mass and gravitational binding energy $E_g=M-E_{M}$. However, definition of the total gravitational energy for the Morris-Thorne wormhole spacetime were stated lated by the Nandi et al. \cite{Nandi:2008ij}:
\begin{equation}
    E_g = M-E_M= \frac{1}{2}\int^{r_2}_{r_0}[1-\sqrt{g_{rr}}]\rho r^2 dr + \frac{r_0}{2}
\end{equation}
Here $r_2$ defines the cut-off scale and $r_0/2$ is the effective gravitational mass.
\begin{figure}[!htbp]
    \centering
    \includegraphics[width=\textwidth]{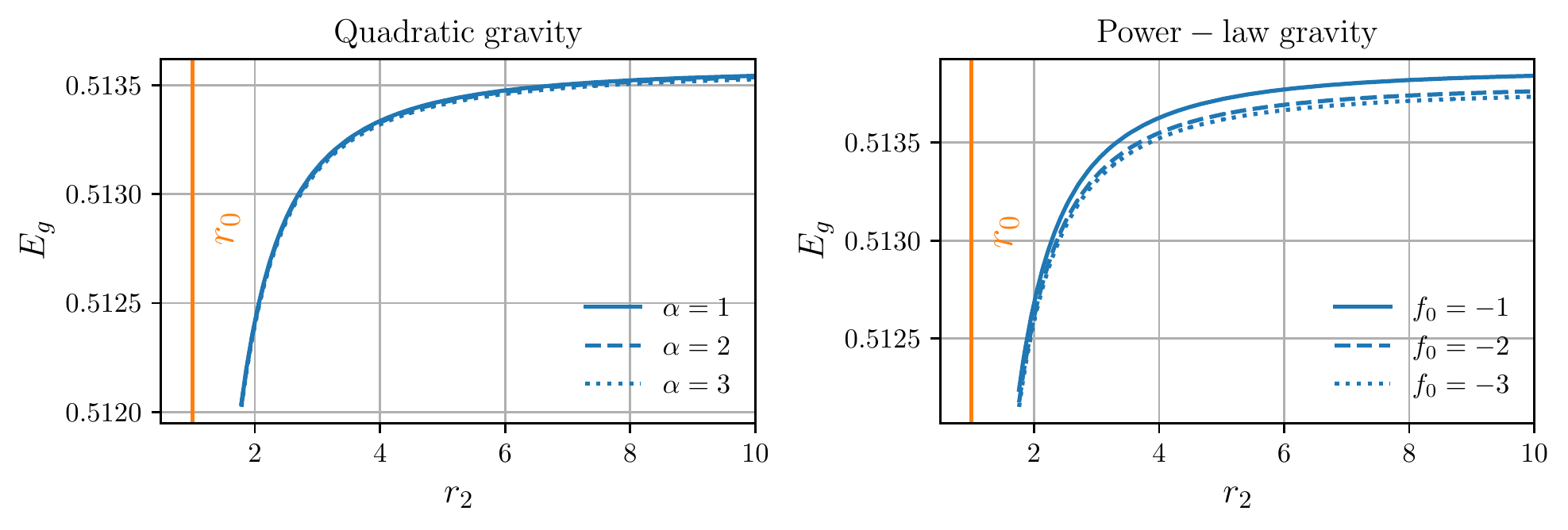}
    \caption{Total gravitational energy for (\textit{first plot}) Quadratic gravity and (\textit{second plot}) power-law gravity with $n=1.1$}
    \label{fig:total}
\end{figure}

As it could be easily remarked, total gravitational energy for both gravity kinds from the Figure (\ref{fig:total}) generally represents the behavior $E_g\approx r_0/2$, and therefore it does approximately equals to the effective gravitational mass. Same behavior for example was observed for Ellis-Bronnikov wormholes within the viable $f(\mathcal{R})$ gravity models (for more details on the subject, see paper \cite{Sokoliuk:2022xcf}).

\section{Conclusions} \label{sec:6}
We presented Casimir spherically symmetric static wormhole solutions for different kinds of $f(\mathcal{R})$ gravity. In our research we used quadratic Starobinsky gravity $f(\mathcal{R})=\mathcal{R}+\alpha \mathcal{R}^2$ and power-law gravity $f(\mathcal{R})=f_0\mathcal{R}^n$. By using Casimir energy density, we derived suitable shape functions for both gravity theories and probed this models via Null, Weak, Dominant and Strong energy conditions.

As well, we investigated the stability of the non-tidal wormholes in the $f(\mathcal{R})$ MOG with the help of modified Tolman- Oppenheimer-Volkoff equation. Furthermore, we found the contribution of the hydrodynaical, anisotropic and extra forces, that arises because of the non-continuity of stress-energy tensor for each shape modified gravity model. For the graphical representation of the forces, present in MTOV, refer to the Figure (\ref{fig:44}). More information about wormhole stability and suitable parameter values could be found in Section (\ref{sec:4}).

Moreover, for the completeness, we as well probe the so-called Volume Integral Quantifier in the current work. It is generally used in order to obtain the exact amount of the exotic matter near the wormhole throat. Numerical solutions for VIQ are respectively placed on the Figure (\ref{fig:VIQ}).

Finally, we derive the total gravitational energy for our wormhole solutions within the both quadratic and power-law $f(\mathcal{R})$ gravities. It was found with the help of numerical integration that equality $E_g\approx r_0/2$, where $r_0/2$ is the effective gravitational mass holds for each wormhole solution.

Now, we think that it will be appropriate to compare our results for ZTF Casimir wormhole with the usual Morris-Thorne wormholes within the same $f(\mathcal{R})$ theories of gravity considered. Morris-Thorne wormholes were investigated in details within the both quadratic and power-law gravities in the work \cite{Sokoliuk:2021rtv}. In the quadratic gravity, it was shown that NEC, SEC for radial and tangential pressures were respected only for $\alpha\geq0$ and only in the case, when $\alpha$ is relatively small (which coincides well with the Casimir wormholes, that was investigated in the present study). However, since authors of the aforementioned paper only used the case with $n=3$ for power-law gravity, it is impossible to compare the results with our near GR case.

In the near future, we plan to investigate the study of the Casimir traversable wormholes with GUP (General Uncertainty Principle) correction in viable $f(\mathcal{R})$ and $f(Q)$ modified gravity models.

\section*{Acknowledgments}
PKS acknowledges National Board for Higher Mathematics (NBHM) under Department of Atomic Energy (DAE), Govt. of India for financial support to carry out the Research project No.: 02011/3/2022 NBHM(R.P.)/R\&D II/2152 Dt.14.02.2022. We are very much grateful to the honorable referees and to the editor for the illuminating suggestions that have significantly improved our work in terms of research quality, and presentation.

\end{document}